\documentclass[aps,pre,reprint]{revtex4-2}

\usepackage{graphicx}
\usepackage{amssymb}
\usepackage{amsmath}
\usepackage{hyperref}

\newcommand{\nn}{\nonumber}
\newcommand{\veps}{\varepsilon}
\newcommand{\tl}{\tilde}

\begin{document}

\title{Electrolubrication in flowing liquid mixtures}

\author{Yoav Tsori}
\email[]{tsori@bgu.ac.il}
\affiliation{Department of Chemical Engineering, Ben-Gurion University of the Negev, Be'er Sheva,
Israel.}

\date{\today}

\begin{abstract}

We describe the ``electrolubrication’’ occurring in liquid mixtures confined 
between two charged surfaces. For a mixture of two liquids, the effective viscosity 
decreases markedly in the presence of a field. The origin of this reduction is field-induced 
phase separation, leading to the formation of two low-viscosity lubrication layers at the 
surfaces. These layers facilitate larger strain at a given stress. The effect is strong if the 
viscosities of the two liquids are sufficiently different, the volume fraction of the less 
viscous liquid is small, the gap between the surfaces is small, and the applied potential is 
large. The phase separation relies on the existence of dissociated ions in the solution. The 
effective viscosity is reduced by a factor $\alpha$; its maximum value is the ratio between 
the viscosities of the two liquids. In most liquids, $\alpha \sim 1$ -- $10$, and in 
mixtures of water and glycerol $\alpha \sim 80$ -- $100$ under relatively small potentials.

\end{abstract}

\maketitle

\section{Introduction}

In the electroviscous effect, the viscosity of particle suspensions depends on the particles' 
charge 
\cite{booth_proc_royal_soc1950,ren_jcis2001,li_colsua2001,sarit_jap2009,ruiz-reina_acis2004,jenner_jcis1995}.
In electrorheological fluids, an external field leads to a large increase in the suspension's viscosity 
\cite{halsey_science1992,hao_adv_mater2001,stangroom_pit1983,gast_acis1989,von_blaaderen_jcp2000}.
In both cases, particle polarization and electrical double-layer buildup lead to long-range interactions 
between the particles that significantly increase the viscosity. The viscoelectric effect refers to changes 
in viscosity due to the effect of external electric fields on molecular dipoles in pure polar liquids. It has 
also been studied extensively in bulk and confined liquids, non-aqueous liquids, and water
\cite{dodd_nature1939,dodd_prslsa1946,dodd_prslsa1951,klein_pnas2022,daiguji_jpcc2017,barisik_langmuir2020,ruckenstein_jcis1981}.

Room-temperature ionic liquids (RTILs) confined between surfaces have been studied extensively as 
potential wear-protective films in recent years. Friction forces are typically measured via an atomic 
force microscope 
\cite{bennewitz_jpcc2019} or surface force balance \cite{perkin_pccp_2012}.
The friction force depends on molecular layering, crowding, and overscreening, often reflecting the 
large size of the molecules \cite{urbakh_naturemat_2022}. In addition, the size-asymmetry between 
anions and cations leads to an asymmetry between the forces measured with positively or negatively 
charged surfaces \cite{atkin_prl_2012}. For moderate surface charge, the friction force between 
surfaces increases with electrode charge for either pure RTILs or RTILs with an added solvent 
\cite{urbakh_acsnano_2017,urbakh_acs_anm_2020}.

This paper describes a new ``electrolubrication'' effect occurring when liquid mixtures are 
confined between two plates. Here, the effective mixture's viscosity can be 
reduced by applying an external potential by $1$ - $3$ orders of magnitude. The effect does not 
depend on specific molecular details and occurs whenever the pure liquids have different viscosities.
In this phenomenon, the external potential leads to field-gradient-induced phase transition 
whereby a thin layer of the more polar solvent (e.g., water) adheres to the surfaces while 
the less polar solvent is in the center of the film \cite{tsori_pnas2007,tsori_pre2021}. As 
shown below, when the polar solvent is the less viscous liquid, the two lubricating layers 
reduce the effective viscosity markedly. The transition voltage depends on the salt 
content, the preferential solvation of ions, temperature, and hydrophobicity/hydrophilicity 
of the plates \cite{tsori_cisc2016}. The phase transition can be of first- or second-order. The 
effective viscosity reduction depends additionally on the ratio between the viscosities of 
the pure liquids, the spacing between the plates, and the relative volume fraction of the 
liquids.

\section{Model} 
\begin{figure}[ht!]
\includegraphics[width=0.49\textwidth,bb=1 560 540 725,clip]{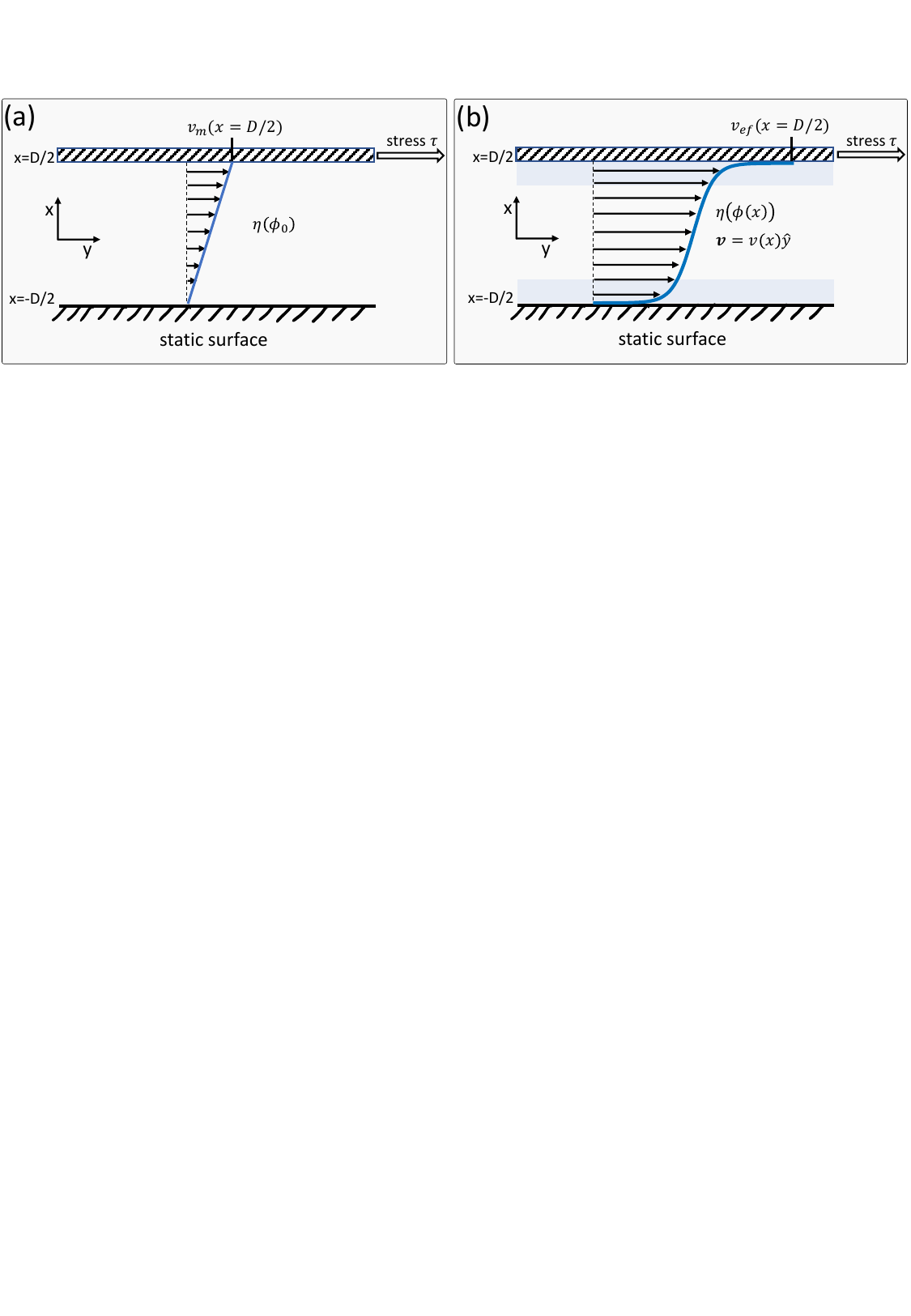}
\caption{Schematic illustration of electrolubrication in a shear experiment. 
A mixture is found in the gap between two parallel and smooth surfaces. The 
bottom one is static and the top one is sheared at a constant stress $\tau$. (a) The 
mixture is homogeneous at composition $\phi_0$. The viscosity $\eta(\phi_0)$ is uniform 
and the velocity is linear in $x$. (b) Under the influence of potential difference across the 
two surfaces, the mixture develops a composition profile leading to inhomogeneities of 
the viscosity $\eta(\phi(x))$. Two lubrication layers are created near the surfaces (shaded in faint blue), 
and the velocity of the top surface is significantly larger than in (a).
}
\label{fig1}
\end{figure}

Consider a mixture of a polar liquid (e.g., water) and a cosolvent placed in between two 
flat and parallel surfaces as in Fig. \ref{fig1}. When the surfaces are charged, ionic 
screening leads to strong field gradients. Ultra-smooth surfaces whose potential can be 
easily adjusted were developed and used even in surface force balance 
experiments, where the distance between surfaces and their smoothness are crucial 
\cite{israelachvili_langmuir2008,tivony_langmuir2015,tivony_langmuir2016,tivony_faraday2017,tivony_jpcc2021,
klein_pnas2022}
For large surface potentials, the ion density at the surfaces becomes very high, and 
therefore we use a modified Poisson-Boltzmann approach, employing the 
incompressibility constraint $\phi+\phi_{\rm cs}+v_0 n^-+v_0 n^+=1$, where $\phi$ 
and $\phi_{\rm cs}$ are the volume fractions of water and cosolvent, respectively, $v_0$ 
is the common molecular volume of all species, and $n^\pm$ is the number density of 
cations and anions. The bulk phase diagram of the mixtures in the 
composition-temperature $(\phi,T)$ plane is assumed to have a critical point at 
$(\phi_c,T_c)$ and a binodal curve $\phi_b(T)$. Liquids with partial miscibility are 
examined at points close to but outside the binodal, while fully miscible liquids are 
modeled at temperatures above $T_c$. We use a simple model where the bulk phase 
diagram in sufficiently low ionic content is symmetric around $\phi_c=1/2$. Thus, if the 
average composition  $\phi=\phi_0$ is in the two-phase region, the equilibrium coexisting 
compositions, $\phi$ and $1-\phi$ are at the binodal.

\subsection{Free Energy} 

In this section, we detail a specific realization of the energy to be used in Sec. \ref{sec_flow_prof} and 
subsequent calculations. The grand potential is
\begin{align}
\label{eq:Omega}
&\Omega[\phi(x),n^\pm(x),\psi(x)]=\int_v\biggl[\frac12 c^2\left(\nabla 
\phi\right)^2
+f_{\rm m}+f_{\rm i}+f_{\rm e} \nn \\ 
&-\lambda^+n^+-\lambda^-n^--\mu\phi/v_0 \biggr]{\rm d}{\bf r} + \int_sf_s{\rm d}{\bf r}_s.
\end{align}
The square-gradient term accounts for the energetic 
cost of composition 
inhomogeneities, where $c^2=\chi/v_0^{1/3}$ is a positive constant. The chemical 
potentials of the cations and anions are $\lambda^\pm$, respectively, and that 
of the mixture is $\mu$. 

The free energy density of mixing is
\begin{align}
\label{eq:fmix}
f_{\rm m}&=\frac{k_BT}{v_0}\left[
\phi\log(\phi)+\phi_{\rm cs}\log(\phi_{\rm cs})+\chi\phi \phi_{\rm cs}\right].
\end{align}
Here, $k_B$ is the Boltzmann constant, $T$ is the absolute temperature, and $\chi\sim1/T$ is the Flory 
interaction parameter. Equation \ref{eq:fmix} leads to an upper critical solution temperature-type phase 
diagram. In the $\phi-T$ plane, a homogeneous phase is stable above the binodal curve, 
$\phi_b(T)$, whereas below it the mixture separates to water-rich and water-poor phases, with 
compositions given by $\phi_b(T)$. The two phases become indistinguishable at the critical 
point $(\phi_c,\chi_c)=(1/2,2)$.

The free energy density of the ions dissolved in the mixture, $f_{\rm i}$, 
is modeled as
\begin{align}
\label{eq:fions}
 f_{\rm i}=&k_BT\bigl[n^+\log (v_0n^+)+n^-\log 
(v_0n^-) \nn \\
&-\phi(\Delta u^+n^+ + \Delta u^-n^-)\bigr],
\end{align}
The first line in Eq. (\ref{eq:fions}) is the entropy of the ions, where $n^\pm$ are their number
densities. The second line models the specific short-range interactions between ions and 
solvents, where the solvation parameters, $\Delta u^\pm$, measure the 
preference of ions toward a local water environment 
\cite{onuki_jcp_2004,onuki_pre_2011,tsori_cisc2016}. For simplicity, we take here 
$\Delta u^+=\Delta u^-$, i.e., both ions are equally hydrophilic.

The electrostatic energy density, $f_{\rm e}$, for a monovalent salt is given by
\begin{align}
\label{eq:fes}
f_{\rm e}=-\frac{1}{2}\veps_0\veps(\phi)(\nabla \psi)^2+e(n^+-n^-)\psi,
\end{align}
where $\psi$ is the electrostatic potential, $e$ is the elementary unit charge and $\veps_0$ is the 
vacuum permittivity. The constitutive relation for $\veps(\phi)$ is 
assumed to depend linearly on the mixture composition: $\veps(\phi)=\veps_{\rm 
cs}+(\veps_w-\veps_{\rm cs})\phi$, where $\veps_w$ and $\veps_{\rm cs}$ are the water and cosolvent 
permittivities, respectively. This $\phi$-dependence of $\veps$ leads, in nonuniform electric fields, to a 
dielectrophoretic force which attracts the high permittivity water toward charged surfaces.

The surface energy density $f_s$ due to the 
contact of the mixture with a solid surface is given by
\begin{equation}
f_s=\Delta\gamma\phi({\bf r}_s),
\label{eq:fsions}
\end{equation}
where ${\bf r}_s$ is a vector on the surface. This expression models the 
short-range interaction between the fluid and the solid. The parameter 
$\Delta\gamma$ measures the difference between the solid--water and 
solid--cosolvent surface tensions. The surfaces can be hydrophilic 
($\Delta \gamma<0$) or hydrophobic $(\Delta \gamma>0)$.

\subsection{Flow Profiles} \label{sec_flow_prof}

The viscosity of the mixture depends on $\phi$, $\phi_{\rm 
cs}$, and $n^\pm$. We denote by $\eta_w$ and $\eta_{\rm cs}$ the 
viscosities of water and cosolvent, respectively. We assume the ionic contribution to the 
viscosity is similar to that of the water, and hence
\begin{eqnarray}\label{eq_const_relation}
\eta=\eta_{\rm cs}\phi_{\rm cs}+\eta_w(1-\phi_{\rm cs}).
\end{eqnarray}

Incompressible binary mixtures in electric fields are governed by the following 
equations \cite{tanaka_jpcm2000,tanaka_jpcm2001,ramos_book,evans_inbook,tritton_book}:
\begin{eqnarray}
\frac{\partial\phi}{\partial t}+{\bf v}\cdot\nabla\phi&=&L_\phi\nabla^2
\frac{\delta f}{\delta\phi},\label{dynam_1st} \\
\nabla\cdot(\veps_0\veps(\phi)\nabla\psi)&=&-e(n^+-n^-)~,\label{dynam_2nd}\\
\nabla \cdot{\bf v}&=&0,\label{dynam_3rd}\\
\rho\left[\frac{\partial {\bf v}}{\partial t}+({\bf v}\cdot{\bf \nabla}){\bf
v}\right]&=&-\nabla p+\nabla\left(\eta\nabla {\bf v}+\eta(\nabla{\bf v})^\dagger\right)\nn\\
&+&{\bf f}_{\rm elec}-\phi\nabla\frac{\delta 
f}{\delta\phi}
\label{dynam_4th},\\
\frac{\partial n^\pm}{\partial t}+\nabla\cdot(n{\bf 
v})&=&L_i\nabla^2\frac{\delta 
f}{\delta n^\pm}. \label{dynam_5th}
\end{eqnarray}
Equation (\ref{dynam_1st}) is a reaction-diffusion equation including advection 
by the flow. ${\bf v}$ is the common velocity of the two liquids. 
$f=f_{\rm m}+f_{\rm i}+f_{\rm e}+(1/2)c^2(\nabla\phi)^2$ is the free energy density of the mixture. 
Equations (\ref{dynam_2nd}) and (\ref{dynam_5th}) are the Poisson-Nernst-Planck equations.
Equation (\ref{dynam_3rd}) is the incompressibility condition and Eq. (\ref{dynam_4th}) is 
the Navier-Stokes equation, with ${\bf f}_{\rm elec}$ from Eq. (\ref{body_surface_force}) below, $p$ the 
pressure,  and a force derivable from gradients of the chemical potential (last term on the right-hand 
side). The liquids' mass density $\rho$ and the Onsager transport coefficients (mobilities) $L_\phi$ and 
$L_i$ are all taken here as constants independent of $\phi$ or $T$.

The body force ${\bf f}_{\rm elec}$ is given by
\begin{eqnarray}\label{body_surface_force}
{\bf f}_{\rm elec}&=&\frac12\nabla\left(\veps_0 E^2 
\phi\frac{\partial \veps}{\partial \phi}\right)_T-\frac12\veps_0E^2\nabla\veps+e(n^+-n^-){\bf E},\nn\\
\end{eqnarray}
where ${\bf E}=-\nabla\psi$ is the electric field. 

{\bf System geometry.} We focus on a simple geometry that allows significant simplification of the
equations. The mixture is held in the gap between two parallel and flat surfaces located at 
$x=\pm D/2$, see Fig. \ref{fig1}. The boundary conditions for ${\bf v}$ are (i) ${\bf v}=0$ at the 
stationary surface at $x=-D/2$, and (ii) constant stress $\tau$ at the moving surface at $x=D/2$. 

In steady-state laminar flow and for infinitely long surfaces in the $y$-direction, it follows that ${\bf 
v}=v(x)\hat{y}$ and all scalar quantities 
$\phi$, $n^\pm$, $\psi$ and $\eta$, depend on $x$ only.  Under these conditions, the complex 
set of equations reduces to a much simpler one: $\phi(x)$, 
$n^\pm(x)$, and $\psi(x)$ obey the equilibrium profiles of a mixture confined by two parallel 
charged plates, and the velocity is given by $\eta(\phi)dv(x)/dx=\tau$.

Therefore, at a given mixture profile $\phi_i(x)$, the flow $v(x)$ is given by
\begin{eqnarray}\label{eq_velocity}
v_i(x)=\tau\int_{x'=-D/2}^x\frac{dx'}{\eta(\phi_i(x'))}.
\end{eqnarray}
The index $i$ is $i=$m for the mixed state and $i=$ef for the mixture in the electric field. 
We are interested in the ``velocity amplification factor'' $\alpha$, defined as the ratio between 
$v_i(x=D/2)$ with and without external potential: $\alpha=v_{\rm ef}(D/2)/v_m(D/2)$. Thus,
\begin{eqnarray}\label{eq_alpha}
\alpha=\frac{1}{D}\int_{x=-D/2}^{x=D/2}\frac{\eta_m}{\eta(\phi_{\rm ef}(x))}dx.
\end{eqnarray}

Without external potential, 
$\phi=\phi_0$ and $\phi_{\rm cs}\approx 1-\phi_0$, provided the volume of ions is sufficiently 
small. The viscosity of this mixed state is $\eta_m= \phi_0\eta_w+(1-\phi_0)\eta_{\rm cs}$. 
The velocity $v_m(x)=(x+D/2)\tau/\eta_m$ is linear in $x$, and $v_m(D/2)=\tau D/\eta_m$. 

$\alpha$ could be large if, in the 
presence of an external field, the less viscous liquid fills the gap and expels the more 
viscous one. To get a significant amplification factor, one needs to have a small value of 
$\phi_0$ (more viscous liquid is a majority), large ratio $\eta_{\rm cs}/\eta_w$, and 
sufficiently large potential to induce the phase separation transition.  

We now make two crude estimates of $v_{\rm ef}(D/2)$ primarily relevant for the case 
where water is a minority phase ($\phi_0<1/2$). In the first one, it is 
assumed that when the field is applied, two pure-water lubrication layers of thickness 
$w$ are formed at the surfaces. The gap between these layers is pure 
cosolvent. In this case
\begin{eqnarray}
v_{\rm ef}(D/2)=\frac{2\tau w}{\eta_w}+\frac{\tau(D-2w)}{\eta_{\rm cs}}.
\end{eqnarray}
Since the velocity of the mixed state is $v_m(D/2)=\tau D/\eta_m$ and $2w/D=\phi_0$ 
we find
\begin{eqnarray}\label{eq_estimate1}
\alpha=\frac{\eta_m}{\eta_{\rm 
cs}}+\phi_0\left(\frac{\eta_m}{\eta_w}-\frac{\eta_m}{\eta_{\rm cs}}\right).
\end{eqnarray}
As a function of $\phi_0$, $\alpha(\phi_0)$ is a parabola with a maximum given at  
$\alpha(\phi_0=1/2)=1/2+(1/4)(\eta_{\rm cs}/\eta_w+\eta_w/\eta_{\rm cs})$. 
It follows that $\alpha \gg 1$ if $\eta_{\rm cs}\gg \eta_w$ and if $\phi_0$ is not close to 
zero or unity. The value of $\alpha$ for a water-glycerol system ($\eta_{\rm cs}\approx 
1412\eta_w$) with $\phi_0=0.2$ yields $\alpha\approx 226$. This figure 
overestimates the value of $\alpha$.

In the second approach, when the external potential is applied, the mixture's composition 
in the gap between the two surfaces is uniform, with $\phi\approx 1-\phi_0$ and $\phi_{\rm 
cs}\approx \phi_0$. The viscosity is then $\eta_{\rm ef}^{\rm 
uni}=(1-\phi_0)\eta_w+\phi_0\eta_{\rm cs}$. The velocity profile is linear in $x$. For the 
water-glycerol system with $\phi_0=0.2$ we find 
\begin{eqnarray}\label{eq_estimate2}
\alpha=\frac{\eta_m}{\eta_{\rm ef}^{\rm uni}}=\frac{\phi_0\eta_w+(1-\phi_0)\eta_{\rm 
cs}}{(1-\phi_0)\eta_w+\phi_0\eta_{\rm cs}}\approx 4.
\end{eqnarray}
This figure underestimates the value of $\alpha$. 

As we will see, the composition in the gap is not uniform--two 
thin nonuniform boundary layers are created near the surfaces. The direct effect of an 
electric double layer in a thin water lubricating film on the viscosity has been evaluated 
elsewhere \cite{meng_tribol_lett2003}. This effect is relatively small compared to the 
electrolubrication discussed here, and consequently, we neglect it.  
In the water-rich layers, the cosolvent content decays to zero with the distance from the 
surfaces $x$. The exact function $\phi(x)$, and consequently $\eta(x)$, is essential in 
determining the velocity profile. In Sec. \ref{results}, we calculate $\phi(x)$ and obtain the 
flow profile.

It is helpful to work with the dimensionless variables defined as
\begin{eqnarray}
\tl{x}&=&\frac{x}{\lambda_D},~~~~~~~~~~~~
\lambda_D^2=\frac{\veps_0\veps(\phi_0)k_BT}{2n_0e^2},\nn\\
\tl{\psi}&=&\frac{e\psi}{k_BT},~~~~~~~~~~~~~~\tl{V}=\frac{eV}{k_BT},\nn\\
\tl{n}^\pm&=&v_0n^\pm,~~~~~~~~~~~~~~\tl{n}_0=v_0n_0,\nn\\
\tl{c}^2&=&c^2\frac{v_0}{\lambda_D^2}\frac{1}{k_BT},~~~~~~~~~
\tl{\mu}=\frac{\mu}{k_BT},\nn\\
\tl{f}_m&=&\frac{v_0f_m}{k_BT},~~~~~~~~~~~~~
\Delta\tl{\gamma}=\frac{\Delta\gamma\lambda_D}{c^2}.
\end{eqnarray}
As we mentioned above, in the parallel-plates geometry, the 
decoupling between the direction perpendicular to the plates ($x$-direction) and parallel 
to them ($y$-direction) means that the quantities $\phi(x)$, $\psi(x)$ and $n^\pm(x)$ 
correspond to their equilibrium profiles. Thus, the ions obey the modified 
Poisson-Boltzmann distribution 
\cite{tsori_cisc2016}
\begin{eqnarray}\label{eqs_npm}
\tl{n}^\pm&=&\frac{P^\pm(1-\phi)}{1+P^++P^-},\\
P^\pm &=&\frac{\tl{n}_0}{1-\phi_0-2\tl{n}_0} 
\exp\left[(\phi-\phi_0)(\chi+\Delta u^\pm)\mp\tl{\psi}\right].\nn
\end{eqnarray}
The two remaining equations, for $\phi$ and $\tilde{\psi}$, are
\begin{eqnarray}\label{eqs_phi_psi}
-\tl{c}^2\tl{\nabla}^2\phi+\frac{\partial 
\tl{f}_m}{\partial\phi}&-&\Delta u^+\tl{n}^+-\Delta u^- 
\tl{n}^-\nn\\
&-&\frac12\frac{2\tl{n}_0}{\veps(\phi_0)}\frac{\partial\veps}
{\partial\phi}(\tl{\nabla} \tl{\psi} )^2-\tl{\mu} =0,\nn\\
\tl{\nabla} \cdot (\veps(\phi)\tl{\nabla}
\tl{\psi})&=&\frac{\veps(\phi_0)}{2\tl{n}_0}(\tl{n}^--\tl{n}^+).
\end{eqnarray}
Their boundary conditions are
\begin{eqnarray}
\frac{d\phi(\tl{x}=\pm 
\tl{D}/2)}{d\tl{x}}&=&\mp\frac{\Delta\tl{\gamma}}{\tl{c}^2},\nn\\
\tl{\psi}(\tl{x}=\pm \tl{D}/2)&=&\pm \frac12 \tl{V}.
\end{eqnarray}

\section{Results}\label{results}
\begin{figure}[ht!]
\includegraphics[width=0.45\textwidth,bb=1 1 440 500,clip]{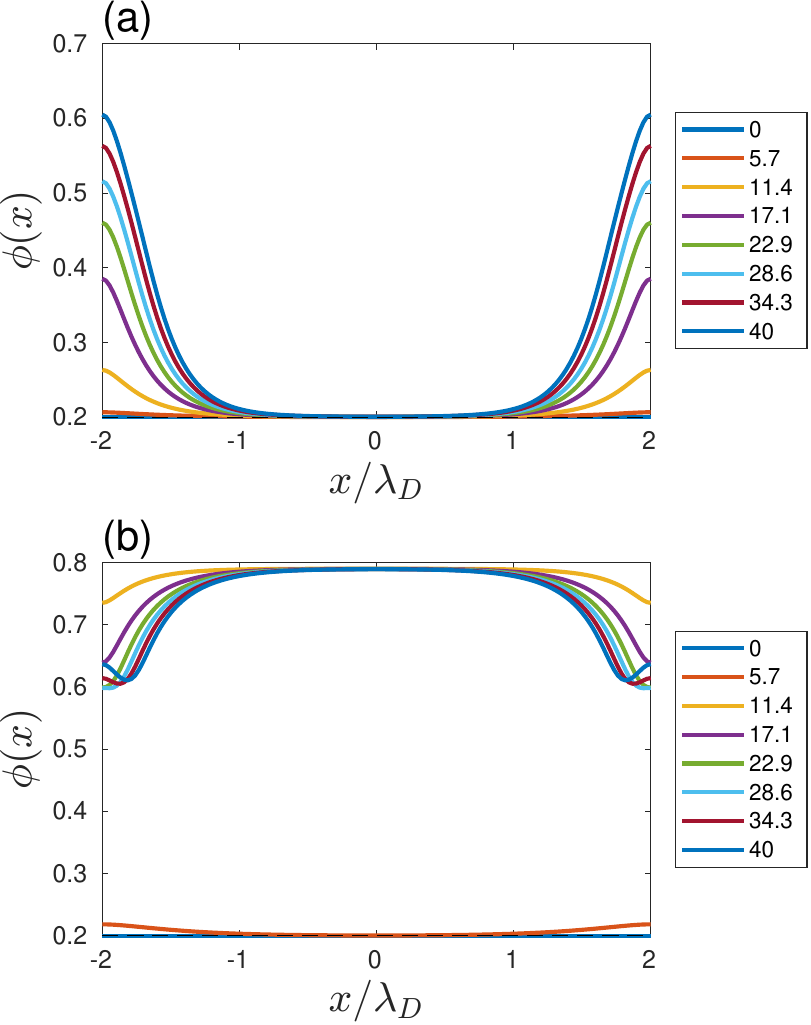}
\caption{Composition profiles for different values of scaled potential $\tilde{V}$ (in 
legend) and fixed bulk 
composition $\phi_0=0.2$. (a) $\chi=0.5$ is much smaller than $\chi_c=2$ ($T$ is well 
above $T_c$). (b) $\chi=\chi_b-0.01$, where $\chi_b=2.31$ is the binodal for $\phi_0$.
Unless stated otherwise, in this and in other figures, we used the parameters of 
water-glycerol mixtures: $\eta_{\rm cs}=\eta_w$, $\veps_{\rm cs}=45$, and 
$\veps_w=80$. In addition $\Delta u^\pm=3$, $\Delta\gamma=0$, $\tilde{n}_0=0.001$, 
$D=3\lambda_D$, and $\tilde{c}=0.4$.
}
\label{fig_profiles_var_Vt}
\end{figure}

We solved Eqs. (\ref{eqs_npm}) and (\ref{eqs_phi_psi}) subject to their boundary conditions. The 
profiles 
$\phi(x)$ are shown in Fig. \ref{fig_profiles_var_Vt} for varying potentials and fixed bulk 
composition $\phi_0=0.2$. In (a) we used $\chi=0.5$, corresponding to temperatures well 
above $T_c$, simulating completely miscible liquids. At vanishing potential, the 
composition is uniform--$\phi(x)=\phi_0$--because the surfaces are taken to be neither 
hydrophobic nor hydrophilic ($\Delta\gamma=0$). As $\tilde{V}$ increases, a profile 
develops. The profiles exhibit a boundary layer at the walls ($\tilde{x}=\pm 1.5$). These 
layers are created due to the electrical double layer and the preferential solvation of ions in 
water. In (b), we used a value of $\chi$ which is close to the bulk binodal $\chi_b(\phi_0)$, 
and increased $\tilde{V}$. While in (a) the profiles $\phi(x)$ evolve gradually with 
$\tilde{V}$, in (b) , they show a discontinuous jump from low $\phi(x)\approx\phi_0$ to high 
$\phi(x)\approx 0.8$ values. This is the signature of the first-order electroprewetting phase transition. At 
$\tilde{V}$ values above the transition, the gap between the surfaces is predominantly 
filled with water. As $\tilde{V}$ further increases, $\phi$ at the surfaces {\it drops} rather 
than increases. This drop is a result of high accumulation of ions at the surfaces; in the 
modified Poisson-Boltzmann framework used here, ions leave less room for water.

\begin{figure}[ht!]
\includegraphics[width=0.45\textwidth,bb=1 1 440 500,clip]{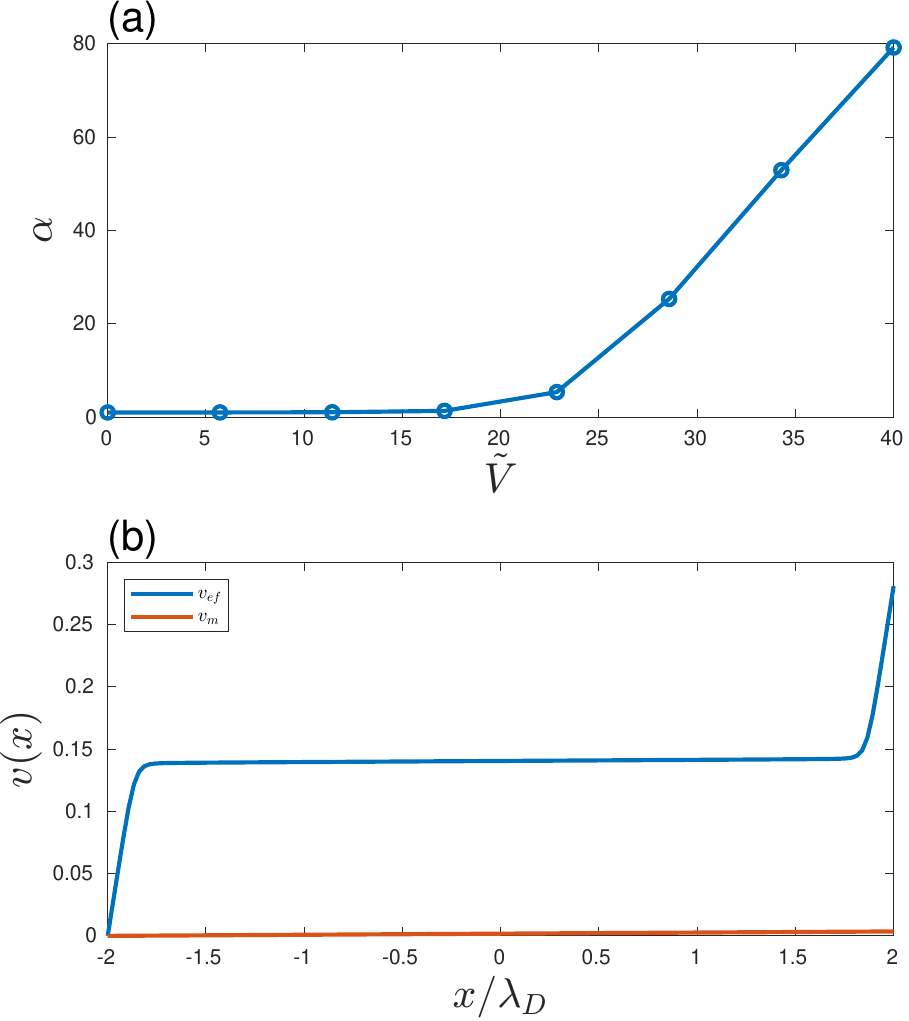}
\caption{Electrolubrication at high temperatures $(\phi_0,\chi)=(0.2,0.5)$. (a) 
$\alpha=v_{\rm ef}/v_m$ from Eq. (\ref{eq_alpha}), the ratio between the 
shear velocities with and without electric field, at the same shear stress, as a function of 
scaled potential $\tilde{V}$. (b) Velocity profiles from Eq. (\ref{eq_velocity}) with 
($v_{\rm ef}$) and without electric field ($v_m$) corresponding to the last point in (a) 
($\tilde{V}=25$). Velocity is in arbitrary units (we used $\tau=1$).
}
\label{fig_velratio1_var_Vt}
\end{figure}
\begin{figure}[ht!]
\includegraphics[width=0.45\textwidth,bb=1 1 440 500,clip]{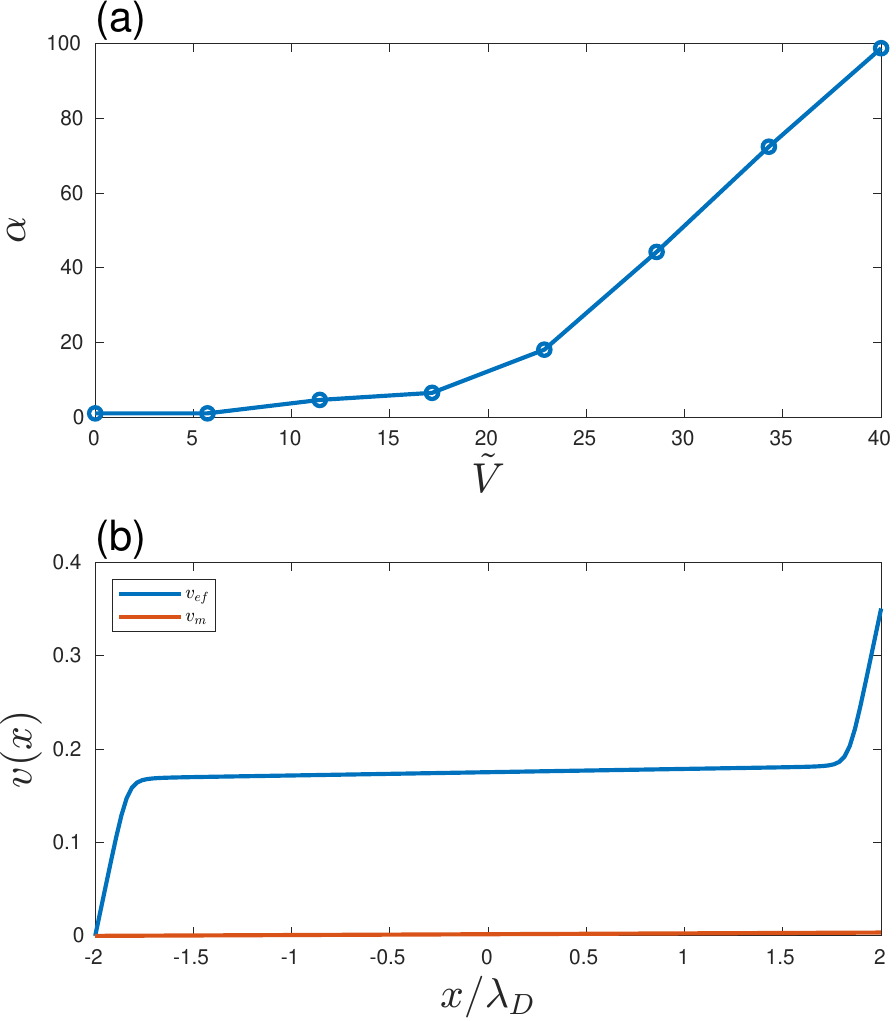}
\rlap{\hspace*{-7cm}\raisebox{6.2cm}{\includegraphics[width=0.18\textwidth,bb=1 1 
440 500,clip]{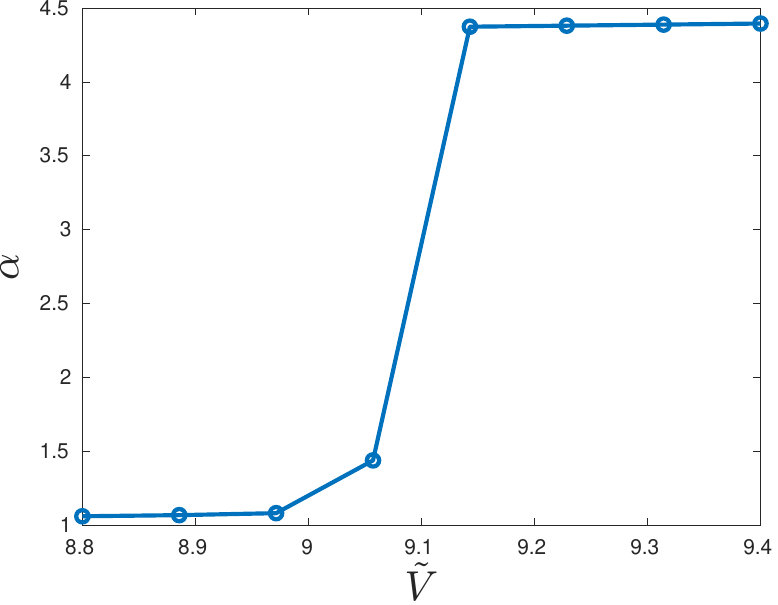}}}
\caption{The same as in Fig. \ref{fig_velratio1_var_Vt} but at low
temperatures: $(\phi_0,\chi)=(0.2,\chi_b-0.01)$ where $\chi_b=2.31$ is the binodal value for 
$\phi_0=0.2$. 
Inset in (a) is a zoom in the range 
$8.8\leq\tilde{V}\leq 9.4$, showing the discontinuity in $\alpha$.
}
\label{fig_velratio2_var_Vt}
\end{figure}

Figures \ref{fig_velratio1_var_Vt} and \ref{fig_velratio2_var_Vt} show the 
electrolubrication effect for the two values of $\chi$ used in Fig. \ref{fig_profiles_var_Vt}. 
In both figures, part (a) is a plot of $\alpha$ from Eq. (\ref{eq_alpha}) vs dimensionless 
potential $\tilde{V}$. The 
maximum value of $\tilde{V}$ here is $40$, corresponding to a potential of about $1$
V. By definition, $\alpha(\tilde{V}=1)=1$. $\alpha(\tilde{V})$ increases with $\tilde{V}$ 
due to the creation of lubrication layers near the surfaces. The thickness of these layers is a 
nontrivial combination of the Debye length $\lambda_D$ and the bulk correlation length 
\cite{tsori_jcp2013}. In Fig. \ref{fig_velratio1_var_Vt}, the 
temperature is high $(\chi=0.5$), $\alpha(\tilde{V})$ is continuous, and reaches a 
value of $14$. In (b) however, the temperature is below $T_c$ and close to the binodal. 
Here $\alpha(\tilde{V})$ grows with $\tilde{V}$ until a critical potential 
$\tilde{V}_c\approx 9.1$, at which it 
jumps discontinuously. Inset shows enlargement of the discontinuity in $\alpha$. 
Above $\tilde{V}_c$, $\alpha(\tilde{V})$ grows continuously. The 
maximum value of $\alpha$, $35$, is higher compared to Fig. \ref{fig_velratio1_var_Vt}.

Parts (b) of Figs. \ref{fig_velratio1_var_Vt} and \ref{fig_velratio2_var_Vt} show the 
velocity profiles calculated by Eq. (\ref{eq_velocity}). Red curves show the classical linear 
shear profile $v_m(x)$ of the homogeneous mixture at the bulk composition $\phi_0$. The 
blue curves show the nonlinear velocity $v_{\rm ef}(x)$ of the inhomogeneous mixture at 
the maximal potential in (a) ($\tilde{V}=25$). The two lubrication layers are evident near 
the surfaces ($\tilde{x}=\pm 1.5$), where the gradient of $v_{\rm ef}$ is large due to the 
small viscosity.

\begin{figure}[ht!]
\includegraphics[width=0.45\textwidth,bb=1 1 440 500,clip]{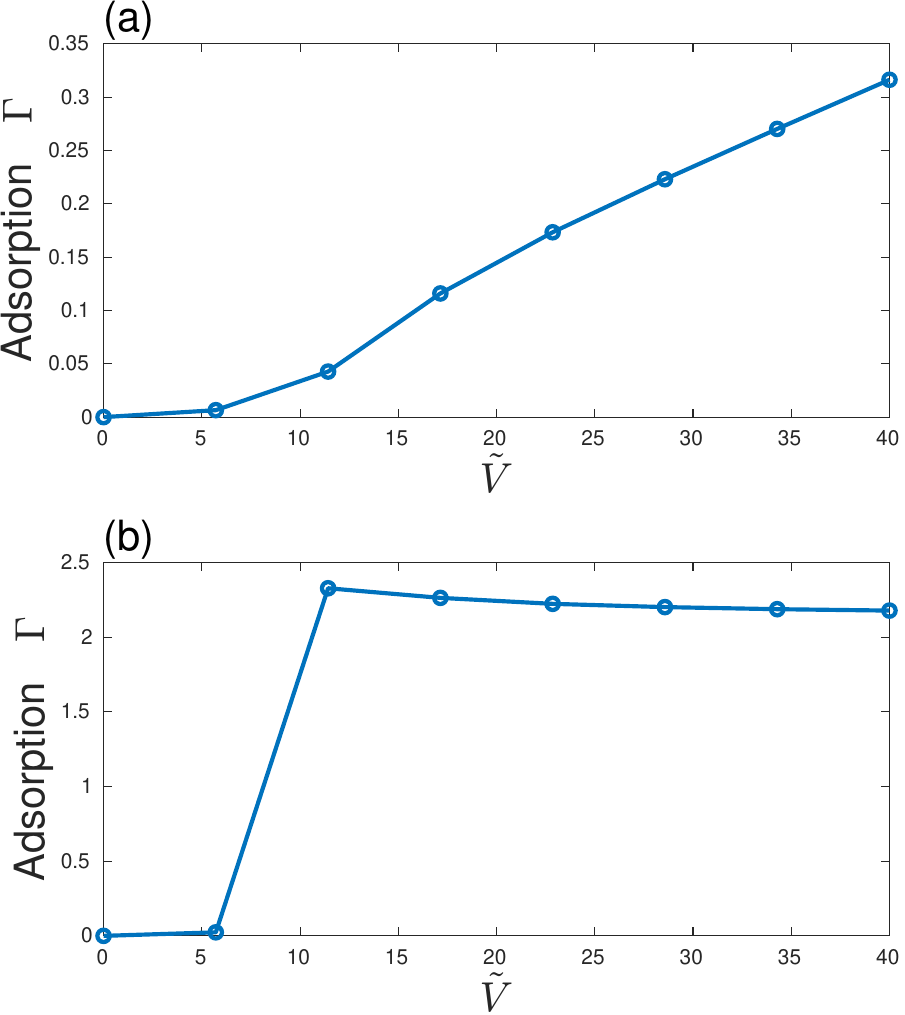}
\caption{Adsorption $\Gamma$ from Eq. (\ref{gamma}) as a function of $\tilde{V}$ for 
the two temperatures $\chi=0.5$ in (a) and $\chi=\chi_b-0.01$ in (b). In both parts 
$\phi_0=0.2$.
}
\label{fig_adsorption_var_Vt}
\end{figure}

In wetting phenomena it is customary to define the adsorption $\Gamma$ as
\begin{eqnarray}\label{gamma}
\Gamma=\int_{-D/2\lambda_D}^{D/2\lambda_D}\left(\phi(\tilde{x})-\phi_0\right)d\tilde{x}.
\end{eqnarray}
$\Gamma$ quantifies the integrated deviation of $\phi(\tilde{x})$ from its bulk value 
$\phi_0$. Figure \ref{fig_adsorption_var_Vt} shows the adsorption vs scaled potential 
$\tilde{V}$ for the low and high temperatures used in previous figures. In (a), $\chi=0.5$ 
and the adsorption grows continuously with $\tilde{V}$. In (b), $\chi=\chi_b-0.01$ is close 
to the binodal. Here $\Gamma(\tilde{V})$ grows continuously for small values of 
$\tilde{V}$. At $\tilde{V}=\tilde{V}_c\approx 9.1$ the first-order phase transition occurs, 
and $\Gamma$ jumps up. This 
discontinuity originates from the jump in profiles seen in Fig. \ref{fig_profiles_var_Vt}(b).
As $\tilde{V}$ further increases, $\Gamma(\tilde{V})$ gradually decreases. As 
mentioned above, at high surface potentials the electric double layer becomes more and more 
crowded, and the reduction in $\phi$ near the surfaces reduces $\Gamma$.
\begin{figure}[ht!]
\includegraphics[width=0.45\textwidth,bb=1 1 440 500,clip]{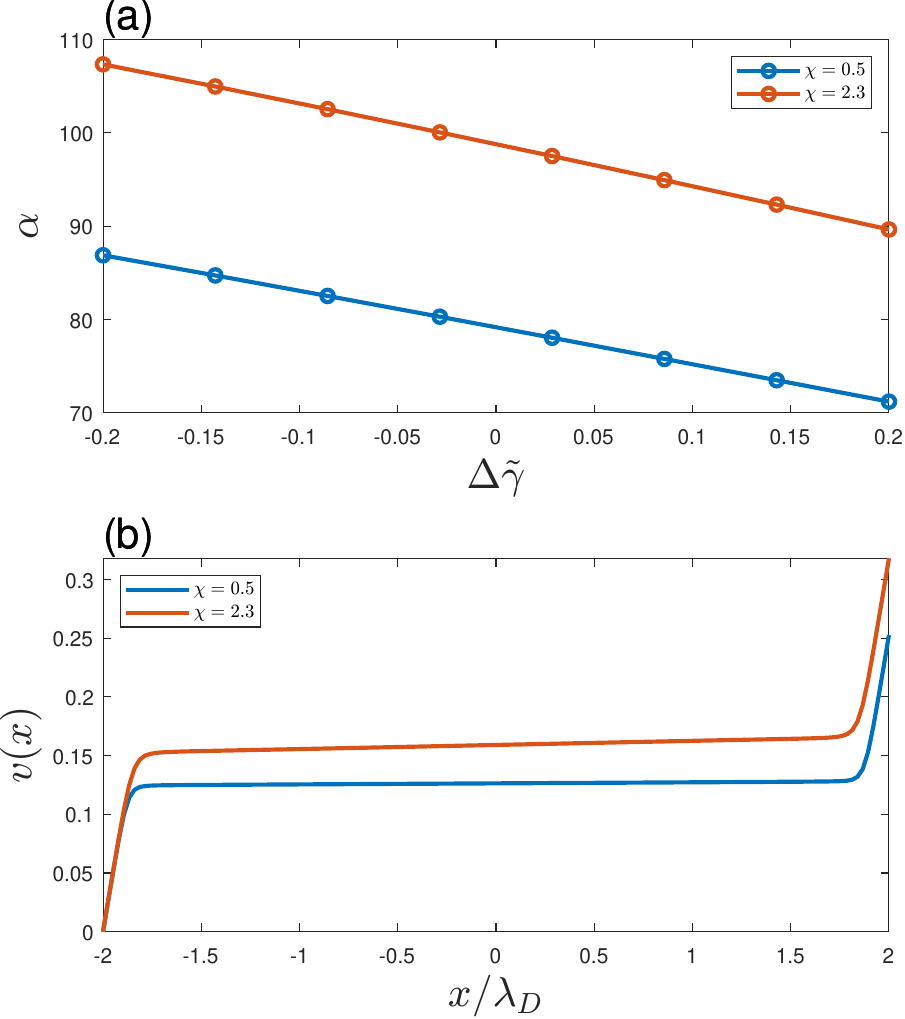}
\caption{The influence of surface hydrophobicity/hydrophilicity on electrolubrication. (a) Flow 
amplification factor $\alpha$ vs $\Delta\tilde{\gamma}$ increasing from negative 
(hydrophilic) to positive (hydrophobic) values. The $\chi$ values (temperatures) are given in the 
legend. $\alpha$ decreases nearly linearly with $\Delta\tilde{\gamma}$. $\alpha$ is larger for low 
temperatures, closer to the binodal. 
(b) The flow profiles $v(x)$ when $\Delta\tilde{\gamma}=0.2$ [two rightmost 
points in (a)]. $v(x)$ curves are larger close to the binodal than far from it.
In both parts, $\phi_0=0.2$ and $\chi_b=2.31$ for this composition. 
}
\label{fig_alpha_var_Dgamma}
\end{figure}

The dependence of electrolubrication on the relative hydrophilicity/hydrophobicity of the surfaces is 
assessed in Fig. \ref{fig_alpha_var_Dgamma}. Here, the value of $\Delta\tilde{\gamma}$ was changed 
between negative (hydrophilic surfaces) to positive values (hydrophobic surfaces) for two different 
temperatures ($\chi$ parameters) as in Figs. \ref{fig_velratio1_var_Vt} and \ref{fig_velratio2_var_Vt}.
In (a), the values of $\alpha$ are shown to decrease nearly linearly with increasing hydrophobicity. 
$\alpha$ is larger when the temperature is closer to the binodal ($\chi=2.3$). Part (b) shows the 
velocity profiles with electric fields at the two temperatures for the rightmost point in (a), i.e., for 
$\Delta\tilde{\gamma}=0.2$). Here again, when the mixture is closer to coexistence (red curve), $v(x)$ 
is larger at any point in the liquid than the high-temperature curve (blue).

\section{Conclusions}

Electrolubrication between two surfaces occurs when an external potential drop leads to 
the layering of confined liquid mixtures. The theory relies on laminar flow, an 
assumption valid for small surface separation $D$. The ratio between the relative 
velocities of the two surfaces with or without field, $\alpha$, can be quite large even at 
small potentials; see vertical axes in Figs. \ref{fig_velratio1_var_Vt} and 
\ref{fig_velratio2_var_Vt}, recalling that $\tilde{V}=40$ corresponds to $1$ V. The 
lubrication effect is predicted for a wide range of temperatures and is strong even for fully 
miscible liquids or far from any coexistence curve [Fig. \ref{fig_velratio1_var_Vt} (a)]. 
Close to a coexistence curve, the electrolubrication jumps discontinuously, as shown in 
the inset of Fig. \ref{fig_velratio2_var_Vt} (a). This electrolubrication is different from the 
``electroactuation'' in RTILs because it predicts a reduction of the effective viscosity even at small 
potentials and is expected to occur even in ordinary liquids. 

Above, we obtained two estimates for $\alpha$, Eqs. (\ref{eq_estimate1}) and 
(\ref{eq_estimate2}). A more realistic expression can be obtained by 
modeling the two layers near the surfaces. Since the electric field decays nearly 
exponentially in the vicinity of the walls we assume that $\eta(x)$ varies as
\begin{eqnarray}
\eta_{\rm ef}&=&\eta_w e^{\beta (x+D/2)}~~~~~ -D/2\leq x\leq -D/2+w,\nn\\
\eta_{\rm ef}&=&\eta_{\rm ef}^{\rm uni}~~~~~~~~~~~~~~~ -D/2+w\leq x\leq 
D/2-w,\nn\\
\eta_{\rm ef}&=&\eta_w e^{\beta (D/2-x)}~~~~~~~~ D/2-w\leq x\leq D/2.
\end{eqnarray}
In this model, the viscosity in the middle of the gap $\eta_{\rm ef}^{\rm uni}$ is uniform. 
What is its value? In Fig.  \ref{fig_profiles_var_Vt}(b)  $\phi\approx 0.8$ and the constitutive 
relation Eq. (\ref{eq_const_relation}) yields $\eta_{\rm ef}^{\rm uni}\approx 285\eta_w$ for 
the water-glycerol system.

The continuity of the viscosity at $x=-D/2+w$ and $x=D/2-w$ sets the value of $\beta$ to 
be $\beta=w^{-1}\log\left(\eta_{\rm ef}^{\rm uni}/\eta_w\right)$.
The velocity of the moving surface can be now readily integrated:
\begin{eqnarray}
v_{\rm ef}(x=D/2)=\frac{2\tau w}{\eta_w}\frac{1-\eta_w/\eta_{\rm 
ef}^{\rm uni}}{\log\left(\frac{\eta_{\rm ef}^{\rm 
uni}}{\eta_w}\right)}+\frac{\tau(D-2w)}{\eta_{\rm 
ef}^{\rm uni}}.
\end{eqnarray}
The ratio between this velocity and $v_m(x=D/2)=\tau D/\eta_m$ is
\begin{eqnarray}
\alpha=2\frac{w\eta_m}{D\eta_w}\frac{1-\eta_w/\eta_{\rm 
ef}^{\rm uni}}{\log\left(\frac{\eta_{\rm ef}^{\rm 
uni}}{\eta_w}\right)}+\frac{(1-2w/D)\eta_m}{\eta_{\rm 
ef}^{\rm uni}}.
\end{eqnarray}
Substitution of the following reasonable figures: $w/D=0.1$, $\phi_0=0.2$, and 
$\eta_{\rm cs}/\eta_w=1412$, yields $\alpha\approx 40$. This value is surprisingly close 
to the full numerical calculation [see, e.g., Fig. \ref{fig_velratio2_var_Vt}(a)] and can be 
important in applications where viscosity control is sought.

An interesting implementation of this effect concerns channel or pipe flow along (say) the 
$y$-direction, driven by a pressure gradient (stationary walls). When a homogeneous 
mixture enters a region in the pipe where the external potential is applied, it will demix in 
the $x$-direction (perpendicular to $y$). The time and length 
scales for this process need to be verified. If the system is translationally invariant along $y$, in 
steady-state, one can still decouple the $x$- and $y$-coordinates as we did here and 
simplify the problem significantly. Another extension of this effect relates to turbulent flow 
or systems lacking apparent symmetry, where layering breaks down and the two 
directions, parallel and perpendicular to the flow, cannot be decoupled.

{\bf Acknowledgment} 
This work was supported by the Israel Science Foundation (ISF) via Grant No. 274/19. 

\bibliography{electrolube_bib}

\end{document}